\documentstyle[prd,aps]{revtex}

\newcommand{\bea}{\begin{eqnarray}}
\newcommand{\eea}{\end{eqnarray}}

\begin{document}

\draft
\twocolumn[\hsize\textwidth\columnwidth\hsize\csname
@twocolumnfalse\endcsname

\title{Cosmological Vorticity in a Gravity with\\
       Quadratic Order Curvature Couplings}
\author{Jai-chan Hwang}
\address{Department of Astronomy and Atmospheric Sciences,
         Kyungpook National University, Taegu, Korea}
\author{Hyerim Noh}
\address{Korea Astronomy Observatory,
         San 36-1, Whaam-dong, Yusung-gu, Daejon, Korea}
\date{\today}
\maketitle

\begin{abstract}

We analyse the evolution of the rotational type cosmological perturbation
in a gravity with general quadratic order gravitational coupling terms.
The result is expressed {\it independently} of the generalized nature of the 
gravity theory, and is simply interpreted as a conservation of the angular 
momentum.

\end{abstract}

\noindent

\vskip2pc]

\section{Introduction}

In a series of work we have been investigating the evolution of cosmological
perturbations in a gravity with quadratic order curvature correction terms
\bea
   S = \int d^4 x \sqrt{-g} \left[ {1 \over 2} \left(
       R + A R^2 + B R^{ab} R_{ab} \right) + L_m \right],
   \label{action}
\eea
where $L_m$ is the matter part Lagrangian.
The gravitational field equation is presented in Eq. (2) of \cite{Rab-GW}.
In contrast with the situation with generalized $f(\phi, R)$ gravity
in Eq. (\ref{action-2}), the gravity with Ricci-curvature square term 
in the action does not have the conformal symmetry to Einstein gravity 
\cite{GGT-1990}.
In \cite{Rab-GW} we analysed the gravitational wave and derived a fourth 
order differential equation.
However, for the background equation we have a symmetry between the
$R^2$ and $R^{ab} R_{ab}$ terms; the effects of the two terms appear
in a common manner in the background equation, see Eq. (\ref{BG-eqs}).

In this paper, we will analyse the vector type perturbation.
In contrast with the gravitational wave studied in \cite{Rab-GW} we will 
find that the Ricci-curvature square term does not add any new feature 
to the evolution of the vector type perturbation: the rotational perturbation
is described by the angular momentum conservation independently of the
generalized nature of the gravity theory.
The main result is Eq. (\ref{AM-conserv}) which is the same equation
valid in Einstein gravity with $A = 0 = B$.
 
Since the analysis for $R^{ab} R_{ab}$ term is new, we present the 
necessary details for the derivation in the Appendices.
We take the same convention as \cite{Rab-GW}.

\section{The rotational type perturbation}

As the background universe model we consider a spatially flat, homogeneous
and isotropic spacetime.
Out of the three (the scalar, vector and tensor) types of general 
perturbations, in this paper, we consider only the transverse vector type 
perturbation. 
To the linear order, due to the symmetry in the background, the three types 
of perturbations decouple from each other and evolve independently.
The tensor type perturbation was considered in \cite{Rab-GW}, 
and the scalar type of perturbation will be considered in future.
The cosmological metric with the general vector type perturbation is
\bea
   ds^2 
   &=& - a^2 d \eta^2 - a^2 B_\alpha d \eta d x^\alpha
   \nonumber \\
   & & 
       + a^2 \left( \delta_{\alpha\beta} + 2 C_{(\alpha,\beta)} \right) 
       d x^\alpha d x^\beta,
   \label{metric}
\eea
where $a(\eta)$ is a cosmic scale factor. 
$B_\alpha ({\bf x}, \eta)$ and $C_\alpha ({\bf x},\eta)$ are transverse
perturbed order variables based on a metric $\delta_{\alpha\beta}$. 
Except for the gauge redundancy, these variables represent the rotational 
type metric perturbations with 
$B^{\alpha}_{\;\;,\alpha} \equiv 0 \equiv C^{\alpha}_{\;\;,\alpha}$. 
The inverse metric, the connection, and the curvatures based on the metric 
in Eq. (\ref{metric}), which are valid to the linear order in $B_\alpha$ 
and $C_\alpha$, are summarized in the Appendices.

Now, we discuss the energy momentum contents supporting the rotational 
perturbation.
The variation of the matter part Lagrangian leads to the energy momentum 
tensor as 
$\delta ( \sqrt{-g} L_m ) \equiv {1 \over 2} \sqrt{-g} T^{ab} \delta g_{ab}$.
We can decompose the energy momentum tensor covariantly as 
\bea
   T_{ab} = \mu u_a u_b + p h_{ab} + q_a u_b + q_b u_a + \pi_{ab},
   \label{Tab}
\eea
where $h_{ab} \equiv g_{ab} + u_a u_b$, 
$q_a u^a \equiv 0 \equiv \pi_{ab} u^b$, and $\pi^a_a \equiv 0$; 
for the covariant set of equations, see \cite{covariant}.
The four vector $u_a$ is decomposed as
\bea
   & & u^0 \equiv a^{-1}, \quad
       u_0 = - a, 
   \nonumber \\
   & & u^\alpha \equiv a^{-1} V^\alpha, \quad
       u_\alpha = a (V_\alpha - B_\alpha).
\eea
We decompose the energy flux $q_a$ as $q_0 = 0$ and 
$q_\alpha \equiv a Q_\alpha$.
$V_\alpha$ and $Q_\alpha$ are based on $\delta_{\alpha\beta}$ with
$V^\alpha_{\;\;\;,\alpha} \equiv 0 \equiv Q^\alpha_{\;\;,\alpha}$.
In the perturbed set of equations, $V_\alpha$ and $Q_\alpha$ always
appear together in a combination $V_\alpha + Q_\alpha / (\mu + p)$.
The energy frame condition $q_\alpha \equiv 0$ corresponds to taking 
$Q_\alpha \equiv 0$, whereas the normal frame condition $u_\alpha \equiv 0$
corresponds to taking $V_\alpha - B_\alpha \equiv 0$, \cite{Israel}.
The combination $V_\alpha + Q_\alpha / (\mu + p)$ counts for the general 
dependence on frames.
The vorticity tensor $\bar \omega_{ab}$ and the shear tensor $\bar \sigma_{ab}$
of the four vector flow $u_a$ are introduced as
\cite{covariant}
\bea
   \bar \omega_{ab} \equiv h_{[a}^c h_{b]}^d u_{c;d}, \quad
       \bar \sigma_{ab} \equiv h_{(a}^c h_{b)}^d u_{c;d} 
       - {1 \over 3} h_{ab} u^c_{\;\; ;c},
\eea
where $t_{[ab]} \equiv {1 \over 2} ( t_{ab} - t_{ba} )$
and $t_{(ab)} \equiv {1 \over 2} ( t_{ab} + t_{ba} )$.
The vorticity tensor $\omega_{ab}$ and the shear tensor $\sigma_{ab}$ based 
on the frame invariant four vector flow 
$u_\alpha^E \equiv u_\alpha + q_\alpha/(\mu + p)$ can be shown as
\bea
   & & \omega_{\alpha\beta} = u^E_{[\alpha,\beta]}
       = a \left( V_{[\alpha,\beta]} + {Q_{[\alpha,\beta]} \over \mu + p}
       - B_{[\alpha,\beta]} \right), 
   \nonumber \\
   & & \sigma_{\alpha\beta} = a \left( V_{(\alpha,\beta)}
       + {Q_{(\alpha,\beta)} \over \mu + p}
       + C_{(\alpha,\beta)}^\prime \right).
   \label{omega-sigma}
\eea
The energy momentum tensor becomes
\bea
   T^0_\alpha = (\mu + p) \left( V_\alpha + {Q_\alpha \over \mu + p}
       - B_\alpha \right), \quad
       \delta T^\alpha_\beta = \Pi^\alpha_\beta,
   \label{Tab-1}
\eea
where we let $\pi^\alpha_\beta \equiv \Pi^\alpha_\beta$ and
consider $\Pi^\alpha_\beta$ as based on $\delta_{\alpha\beta}$.
$\mu$ and $p$ are the energy density and the pressure based on $u_a$ as
\bea
   \mu \equiv T_{ab} u^a u^b = - T^0_0, \quad
       p \equiv {1 \over 3} T_{ab} h^{ab} = {1 \over 3} T^\alpha_\alpha.
\eea 
To the linear order, $\mu$ and $p$ are frame invariant, \cite{Israel}.

\section{The gauge issue}

Under the gauge transformation $\hat x^a = x^a + \tilde \xi^a (x^e)$ the metric
transforms as
\bea
   \delta \hat g_{ab} = \delta g_{ab} - g_{ab,c} \tilde \xi^c
       - g_{cb} \tilde \xi^c_{\;\;,a} - g_{ac} \tilde \xi^c_{\;\;,b},
\eea
and similarly for the energy momentum tensor $T_{ab}$.
We consider the vector type transformation, thus $\tilde \xi^0 \equiv 0$ and 
$\tilde \xi^\alpha \equiv \xi^\alpha$ with $\xi^\alpha$ based on 
$\delta_{\alpha\beta}$ and $\xi^\alpha_{\;\;,\alpha} \equiv 0$.
{}From the gauge transformation properties of the metric and the energy 
momentum tensor we can show that
\bea
   & & \hat B_\alpha = B_\alpha + \xi_\alpha^\prime, \quad
       \hat C_\alpha = C_\alpha - \xi_\alpha,
   \nonumber \\
   & & \hat V_\alpha + {\hat Q_\alpha \over \mu + p}
       = V_\alpha + {Q_\alpha \over \mu + p} + \xi_\alpha^\prime,
\eea
and $\Pi^\alpha_\beta$ is gauge invariant.
Thus, we have three different gauge conditions: the condition 
$B_\alpha \equiv 0$ or $V_\alpha + Q_\alpha / (\mu + p) \equiv 0$ does not 
fix the gauge condition completely, whereas, the $C_\alpha \equiv 0$ 
condition completely fixes the gauge condition.
Thus, without losing generality we can take $C_\alpha \equiv 0$ as a 
preferred gauge condition; we call this the C-gauge condition, \cite{Aniso}.
Under the C-gauge condition, $B_\alpha$ and $V_\alpha + Q_\alpha/(\mu + p)$ 
are {\it the same as} the following gauge invariant combinations, respectively
\bea
   B_\alpha + C_\alpha^\prime, \quad
       V_\alpha + {Q_\alpha \over \mu + p} + C_\alpha^\prime.
\eea
We have one more gauge invariant combination 
\bea
   V_\alpha + {Q_\alpha \over \mu + p} - B_\alpha.
\eea

\section{Harmonic decomposition}

We treat the spatial dependence of the perturbed variables using the vector 
type harmonic function $Y_\alpha$ introduced as \cite{Bardeen}
\bea
   \Delta Y_\alpha \equiv - k^2 Y_\alpha, \quad
       Y_{\alpha\beta} \equiv - { 1\over k} Y_{(\alpha,\beta)}, \quad
       Y^\alpha_{\;\;\;,\alpha} \equiv 0.
   \label{Y}
\eea
We decompose the perturbation variables as
\bea
   & & B_\alpha ({\bf x}, t) \equiv b (t) Y_\alpha ({\bf x}), \quad
       C_\alpha \equiv c Y_\alpha, 
   \nonumber \\
   & & V_\alpha \equiv v Y_\alpha, \quad
       Q_\alpha \equiv q Y_\alpha, \quad
       \Pi^\alpha_\beta \equiv p \pi_T Y^\alpha_\beta.
   \label{decomposition}
\eea
We introduce the following {\it gauge invariant} variables:
\bea
   & & V_\alpha + {Q_\alpha \over \mu + p} - B_\alpha
       = \left( v + {q \over \mu + p} - b \right) Y_\alpha
       \equiv v_\omega Y_\alpha,
   \nonumber \\
   & & V_\alpha + {Q_\alpha \over \mu + p} + C_\alpha^\prime
       = \left( v + {q \over \mu + p} + c^\prime \right) Y_\alpha
       \equiv v_\sigma Y_\alpha,
   \nonumber \\
   & & B_\alpha + C_\alpha^\prime = ( v_\sigma - v_\omega ) Y_\alpha 
       \equiv \Psi Y_\alpha.
   \label{Psi}
\eea 
{}From Eq. (\ref{omega-sigma}) we have
\bea
   \omega_{\alpha\beta} = a v_\omega Y_{[\alpha,\beta]}, \quad
       \sigma_{\alpha\beta} = a v_\sigma Y_{(\alpha,\beta)}.
\eea
The amplitude square of the vorticity and the shear are defined as
\bea
   \omega^2 \equiv {1 \over 2} \omega^{ab} \omega_{ab}, \quad
       \sigma^2 \equiv {1 \over 2} \sigma^{ab} \sigma_{ab}.
\eea
We can show 
\bea
   & & \omega = \left| {1 \over a} v_\omega \right|
       \sqrt{ {1 \over 2} Y^{\alpha|\beta} Y_{[\alpha,\beta]} }, 
   \nonumber \\
   & & \sigma = \left| {1 \over a} v_\sigma \right|
       \sqrt{ {1 \over 2} Y^{\alpha|\beta} Y_{(\alpha,\beta)} }.
\eea
Thus,
\bea
   v_\omega \propto a \omega, \quad
       v_\sigma \propto a \sigma,
\eea
and we can interprete $v_\omega$ and $v_\sigma$ as the velocity variables
related to the vorticity and the shear, respectively; 
in \cite{Bardeen,GGT-1990,PRW} $v_\omega$ and $v_\sigma$ are written 
as $v_c$ and $v_s$, respectively.

\section{Equations for the cosmological vorticity}

Using the quantities presented in the Appendix B (thus taking the C-gauge)
and Eq. (\ref{Tab-1}), the gravitational field equation in Eq. (2) of 
\cite{Rab-GW} becomes:
\bea
   \delta T^0_\alpha
   &=& (1 + 2 A R) R^0_\alpha
   \nonumber \\
   & & - B \left[ \ddot R^0_\alpha + 5 H \dot R^0_\alpha
       - \left( 2 \dot H + 2 H^2 + {\Delta \over a^2} \right) 
       R^0_\alpha \right]
   \nonumber \\
   &=& ( \mu + p) \left( V_\alpha + {Q_\alpha \over \mu + p} 
       - B_\alpha \right),
   \label{Eq1} \\
   \delta T^\alpha_\beta
   &=& {1 \over 2 a^3} \left[ ( 1 + 2 A R ) a^2  \left( B^\alpha_{\;\;,\beta}
       + B_\beta^{\;\;|\alpha} \right) \right]^\cdot
   \nonumber \\
   & & - B \Bigg[ \delta \ddot R^\alpha_\beta + 3 H \delta \dot R^\alpha_\beta
       - \left( 6 H^2 + 6 \dot H + {\Delta \over a^2} \right)
       \delta R^\alpha_\beta
   \nonumber \\
   & & + \left( - 6 H \dot H - 3 \ddot H + H {\Delta \over a^2} \right)
       {1 \over a} \left( B^\alpha_{\;\;,\beta}
       + B_\beta^{\;\;|\alpha} \right) \Bigg]
   \nonumber \\
   &=& \Pi^\alpha_\beta,
   \label{Eq2}
\eea
where $R = 6 ( \dot H + 2 H^2 )$.
The equations for the background are derived in Eq. (7) of \cite{Rab-GW}:
\bea
   & & H^2 + 2 \left( 3 A + B \right) \left( 2 H \ddot H - \dot H^2 
       + 6 H^2 \dot H \right) = {1 \over 3} \mu, 
   \nonumber \\
   & & \dot \mu = - 3H \left( \mu + p \right).
   \label{BG-eqs}
\eea
{}From Eqs. (\ref{Rab},\ref{Psi},\ref{Y},\ref{Tab-1},\ref{decomposition}) 
we have
\bea
   & & R^0_\alpha = {k^2 \over 2 a^2} \Psi Y_\alpha, \quad
       \delta R^\alpha_\beta = - {k \over a^3} \left( a^2 \Psi \right)^\cdot
       Y^\alpha_\beta,
   \\
   & & T^0_\alpha = \left( \mu + p \right) v_\omega Y_\alpha, \quad
       \delta T^\alpha_\beta = p \pi_T Y^\alpha_\beta,
\eea
and Eqs. (\ref{Eq1},\ref{Eq2}) can be combined to give
\bea
   & & {k^2 \over 2 a^2} \left\{ \left( 1 + 2 A R \right) \Psi
       - B \left[ \ddot \Psi + H \dot \Psi - 
       \left( {2 \over 3} R + {\Delta \over a^2} \right) \Psi \right] \right\}
   \nonumber \\
   & & \qquad
       = \left( \mu + p \right) v_\omega, 
   \label{Psi-v} \\
   & & {1 \over a^4} \Big[ a^4 \left( \mu + p \right) v_\omega \Big]^\cdot 
       = - {k \over 2 a} p \pi_T.
   \label{AM-conserv}
\eea
Equation (\ref{AM-conserv}) implies the angular momentum conservation:
for $p \pi_T = 0$ we have `angular momentum'
$\sim a^3 (\mu + p) \times a \times v_\omega \sim$ constant in time.
Notice that, remarkably, the angular momentum conservation relation in
Eq. (\ref{AM-conserv}) involves neither $A$ nor $B$ in Eq. (\ref{action}), 
thus, does not depend on the character of the generalized nature of the 
gravity theories.

\section{Discussions}

The cosmological evolution of the rotational perturbation in Einstein gravity
and its simple characterization by the angular momentum conservation were
first presented in a classic study by Lifshitz in \cite{Lifshitz}.
The extension of the result to the generalized gravity with an action
\bea
   & & S = \int d^4 x \sqrt{-g} \Bigg[ {1 \over 2} f(\phi, R) 
       - {1 \over 2} \omega(\phi) \phi^{;a} \phi_{,a} - V(\phi) 
   \nonumber \\
   & & \qquad \qquad \qquad \qquad
       + \; L_m \Bigg],
   \label{action-2}
\eea
was presented in \cite{GGT-1990,PRW}; $\phi$ is a dilaton field.
In this case, instead of Eq. (\ref{Psi-v}), we have
\bea
   {k^2 \over 2 a^2} F \Psi = \left( \mu + p \right) v_\omega,
   \label{Psi-v-2}
\eea
where $F \equiv \partial f / (\partial R)$,
and Eq. (\ref{AM-conserv}) remains intact.
{}For the gravity in Eq. (\ref{action}) with $B = 0$, we have $F = 1 + 2 A R$.
In \cite{Nariai-1,Nariai-2} the authors considered instabilities of the 
homogeneous and isotropic world model based on $L \sim f(R)$ gravity 
under the gravitational wave and rotational perturbations.
The results of both the gravitational wave and rotational perturbations 
agree with \cite{GGT-1990,Rab-GW} and present work, respectively.
In \cite{Nariai-1} the author analysed the scalar type perturbations
in $L \sim R^n$ gravity using the synchronous gauge;
recent progress in \cite{HN-GGT} successfully resolved the issue 
in a simple and unified manner in the gravity theories in Eq. (\ref{action-2})
using the uniform-curvature gauge.
[The referee has informed us the important works in \cite{Nariai-1,Nariai-2}.]

The equation for a multi-component situation in $L_m$ in the context of
Eq. (\ref{action-2}) is presented in Sec. 5.1 of \cite{PRW}.
The evolution of background universe follows Eq. (\ref{BG-eqs}) where
the fluid quantities, $\mu$ and $p$, concern the matter part Lagrangian.
{}For $p \equiv {\rm w} \mu$ with ${\rm w} = {\rm constant}$
we have $\mu \propto a^{-3 (1 + {\rm w})}$.
Thus, for vanishing anisotropic pressure of the matter part,
$p \pi_T = 0$, we have
\bea
   v_\omega \propto a^{-1 + 3 {\rm w}},
\eea
thus, $v_\omega \propto a^0, \; a^{-1}, \; a^{-4}$ for 
the radiation flow (${\rm w} = {1 \over 3}$), the matter flow (${\rm w} = 0$),
and the inflationary matter flow (${\rm w} \simeq -1$), respectively.
This result concerning the vorticity does not depend on the generalized nature 
of the gravity theories of the types in Eqs. (\ref{action},\ref{action-2}).
The related disturbances in the metric ($\Psi$) and the corresponding shear 
of the flow ($v_\sigma$) do, however, depend on the generalized nature of the 
gravity as in Eqs. (\ref{Psi-v},\ref{Psi-v-2}).

\section*{Acknowledgments}

This work was supported by the KOSEF, Grant No. 95-0702-04-01-3  
and through the SRC program of SNU-CTP.

\vskip .5cm
\centerline{\bf APPENDIX A: LINEAR ORDER QUANTITIES}

\vskip .5cm
\noindent
The inverse metric ($0 = \eta$):
\bea
   & & g^{00} = - {1 \over a^2}, \quad
       g^{0\alpha} = - {1 \over a^2} B^{\alpha}, 
   \nonumber \\
   & & g^{\alpha\beta} = {1 \over a^2} \left( \delta^{\alpha\beta}
       - 2 C^{(\alpha|\beta)} \right).
\eea
The connection:
\bea
   & & \Gamma^0_{00} = {a^\prime \over a}, \quad
       \Gamma^0_{0\alpha} = - {a^\prime \over a} B_\alpha, \quad
       \Gamma^\alpha_{00} = - B^{\alpha\prime} - {a^\prime \over a} B^\alpha,
   \nonumber \\
   & & \Gamma^0_{\alpha\beta} = {a^\prime \over a} \delta_{\alpha\beta}
       + B_{(\alpha,\beta)} + C^\prime_{(\alpha,\beta)} 
       + 2 {a^\prime \over a} C_{(\alpha,\beta)}, 
   \nonumber \\
   & & \Gamma^\alpha_{0\beta} 
       = {a^\prime \over a} \delta^\alpha_\beta
       - {1 \over 2} \left( B^\alpha_{\;\;,\beta}
       - B_\beta^{\;\; |\alpha} \right)
       + {1 \over 2} \left( C^\alpha_{\;\;,\beta}
       + C_\beta^{\;\; |\alpha} \right)^\prime,
   \nonumber \\
   & & \Gamma^\alpha_{\beta\gamma} 
       = {a^\prime \over a} B^{\alpha} \delta_{\beta\gamma}
       + C^{\alpha}_{\;\;,\beta\gamma}.
\eea
Riemann curvature:
\bea
   & & R^a_{\;\;b00} = 0 = R^0_{\;\;0\alpha\beta}, \quad
       R^0_{\;\;00\alpha} 
       = - \left( {a^\prime \over a} \right)^\prime B_{\alpha},
   \nonumber \\
   & & R^0_{\;\;\alpha 0 \beta} 
       = \left( {a^\prime \over a} \right)^\prime \delta_{\alpha\beta} 
       + B^\prime _{(\alpha,\beta)} + {a^\prime \over a} B_{(\alpha,\beta)}
   \nonumber \\
   & & \qquad
       + C^{\prime\prime}_{(\alpha,\beta)}
       + {a^\prime \over a} C^\prime_{(\alpha,\beta)}
       + 2 \left( {a^\prime \over a} \right)^\prime C_{(\alpha,\beta)},
   \nonumber \\
   & & R^0_{\;\;\alpha\beta\gamma} = B_{[\gamma,\beta]\alpha}
       + C^\prime_{[\gamma,\beta]\alpha},
   \nonumber \\
   & & R^\alpha_{\;\; 00\beta} 
       = \left( {a^\prime \over a} \right)^\prime \delta^\alpha_\beta 
   \nonumber \\
   & & \qquad
       + {1 \over 2} \left( B^\alpha_{\;\;,\beta}
       + B_\beta^{\;\;|\alpha} \right)^{\prime}
       + {1 \over 2} {a^\prime \over a} \left( B^\alpha_{\;\;,\beta}
       + B_\beta^{\;\;|\alpha} \right)
   \nonumber \\
   & & \qquad
       + {1 \over 2} \left( C^\alpha_{\;\;,\beta}
       + C_\beta^{\;\;|\alpha} \right)^{\prime\prime}
       + {1 \over 2} {a^\prime \over a} \left( C^\alpha_{\;\;,\beta}
       + C_\beta^{\;\;|\alpha} \right)^{\prime},
   \nonumber \\
   & & R^\alpha_{\;\; 0 \beta\gamma} 
       = B^{\;\;\;\;\;\;\; |\alpha}_{[\gamma,\beta]}
       + 2 \left( {a^\prime \over a} \right)^2
       \delta^\alpha_{[\gamma} B_{\beta]}
       + C^{\prime \;\;\;\;\;\; |\alpha}_{[\gamma,\beta]},
   \nonumber \\
   & & R^\alpha_{\;\; \beta 0\gamma} 
       = \left( {a^\prime \over a} \right)^\prime B^\alpha \delta_{\beta\gamma}
       + \left( {a^\prime \over a} \right)^2
       \left( - B^\alpha \delta_{\beta\gamma}
       + B_\beta \delta^\alpha_\gamma \right)
   \nonumber \\
   & & \qquad
       + {1 \over 2} \left( B^\alpha_{\;\;,\beta}
       - B_\beta^{\;\;|\alpha} \right)_{,\gamma}
       + {1 \over 2} \left( C^\alpha_{\;\;,\beta}
       - C_\beta^{\;\;|\alpha} \right)^\prime_{,\gamma},
   \nonumber \\
   & & R^\alpha_{\;\;\beta\gamma\delta}
       = 2 \left( {a^\prime \over a} \right)^2 
       \delta^\alpha_{[\gamma} \delta_{\delta]\beta}
   \nonumber \\
   & & \qquad
       + {a^\prime \over a} \left( 
       \delta^\alpha_{[\gamma} B_{\delta],\beta}
       + B_{\beta,[\delta} \delta^\alpha_{\gamma]}
       + \delta^\alpha_{[\gamma} C^\prime_{\delta],\beta}
       + C^\prime_{\beta,[\delta} \delta^\alpha_{\gamma]} \right)
   \nonumber \\
   & & \qquad
       + 2 \left( {a^\prime \over a} \right)^2 \left(
       \delta^\alpha_{[\gamma} C_{\delta],\beta}
       + C_{\beta,[\delta} \delta^\alpha_{\gamma]} \right)
   \nonumber \\
   & & \qquad 
       + {a^\prime \over a} \delta_{\beta[\delta} \left[
       B^\alpha_{\;\;,\gamma]} + B_{\gamma]}^{\;\;\;|\alpha} 
       + \left( C^\alpha_{\;\;,\gamma]} 
       + C_{\gamma]}^{\;\;\;|\alpha} \right)^\prime \right].
\eea
In the following and the Appendix B we present only the perturbed parts 
and ignore the (0,0) components which give null results to the perturbed order;
the results valid in the background order are presented in the Appendices
of \cite{Rab-GW}.

\noindent
Ricci curvature: $\delta R^0_0 = 0$,
\bea
   & & R^0_\alpha = - {1 \over 2} {\Delta \over a^2} \left( B_\alpha 
       + C^\prime_\alpha \right),
   \nonumber \\
   & & \delta R^\alpha_\beta
       = {1 \over a^2} \Bigg\{ 
       {1 \over 2} \left( B^\alpha_{\;\;,\beta}
       + B_\beta^{\;\;|\alpha} \right)^\prime
       + {a^\prime \over a} \left( B^\alpha_{\;\;,\beta}
       + B_\beta^{\;\;|\alpha} \right)
   \nonumber \\
   & & \qquad
       + {1 \over 2} \left( C^\alpha_{\;\;,\beta}
       + C_\beta^{\;\;|\alpha} \right)^{\prime\prime}
       + {a^\prime \over a} \left( C^\alpha_{\;\;,\beta}
       + C_\beta^{\;\;|\alpha} \right)^\prime \Bigg\}.
\eea
Scalar curvature: $\delta R = 0$.

\noindent
Curvature combinations:
\bea
   & & \delta ( R^a_b R^b_a ) = 0 = \delta ( R^{cd} R^0_{\;\; cd 0} ),
   \nonumber \\
   & & R^{cd} R^0_{\;\; cd \alpha} 
       = \left( {a^\prime \over a} \right)^2 {\Delta \over a^4}
       \left( B_\alpha + C^\prime_\alpha \right),
   \nonumber \\
   & & \delta ( R^{cd} R^\alpha_{\;\; cd \beta} )
       = - {2 \over a^4} {a^\prime \over a}
       \left( {a^\prime \over a} \right)^\prime 
   \nonumber \\
   & & \qquad \times
       \left[ B^\alpha_{\;\;,\beta} + B_\beta^{\;\;|\alpha}
       + \left( C^\alpha_{\;\;,\beta} + C_\beta^{\;\;|\alpha} 
       \right)^{\prime} \right]
   \nonumber \\
   & & \qquad 
       + {1 \over 2 a^4} \left[ - 3 \left( {a^\prime \over a} \right)^{\prime}
       + \left( {a^\prime \over a} \right)^2 \right] 
   \nonumber \\
   & & \qquad \times
       \left[ 
       \left( B^\alpha_{\;\;,\beta} + B_\beta^{\;\;|\alpha} \right)^\prime
       + \left( C^\alpha_{\;\;,\beta} + C_\beta^{\;\;|\alpha} 
       \right)^{\prime\prime} \right].
\eea
Covariant derivatives: 
\bea
   & & \delta ( R^{;0}_{\;\;\; 0} ) = 0 = R^{;0}_{\;\;\;\alpha},
   \nonumber \\
   & & \delta ( R^{;\alpha}_{\;\;\;\beta} )
       = - {1 \over 2 a^2} R^\prime \left[
       B^\alpha_{\;\;,\beta} + B_\beta^{\;\;|\alpha}
       + \left( C^{\alpha}_{\;\;,\beta} 
       + C_\beta^{\;\;|\alpha} \right)^\prime \right],
   \nonumber \\
   & & \Box R = 0 = \Box R^0_0,
   \nonumber \\
   & & \Box R^0_\alpha = {1 \over a^2} \Bigg\{ - R^{0\prime\prime}_\alpha 
       - 4 {a^\prime \over a} R^{0\prime}_\alpha
   \nonumber \\
   & & \qquad
       + \left[ \Delta + 2 \left( {a^\prime \over a} \right)^\prime
       - 4 \left( {a^\prime \over a} \right)^2 \right] R^0_\alpha \Bigg\},
   \nonumber \\
   & & \delta ( \Box R^\alpha_\beta )
   \nonumber \\
   & & \qquad
       = {1 \over a^2} \Bigg\{ - \delta R^{\alpha\prime\prime}_\beta
       - 2 {a^\prime \over a} \delta R^{\alpha\prime}_\beta
       + \left[ \Delta + 2 \left( {a^\prime \over a} \right)^2 \right] 
       \delta R^\alpha_\beta
   \nonumber \\
   & & \qquad
       - {a^\prime \over a^3} \left[ \Delta 
       + 4 \left( {a^\prime \over a} \right)^\prime
       - 4 \left( {a^\prime \over a} \right)^2 \right]
   \nonumber \\
   & & \qquad \times
       \left[ B^\alpha_{\;\;,\beta} + B_\beta^{\;\;|\alpha}
       + \left( C^\alpha_{\;\;,\beta} + C_\beta^{\;\;|\alpha} \right)^\prime
       \right] \Bigg\}.
\eea

\vskip .5cm
\centerline{\bf APPENDIX B: IN TERMS OF $t$}

\vskip .5cm
Using $t$ as the time variable ($dt = a d \eta$), and taking
$C_\alpha \equiv 0$ as the gauge condition, we have ($H \equiv \dot a/a$):
\bea
   & & R^0_\alpha = - {1 \over 2} {\Delta \over a^2} B_\alpha, \;\;
       \delta R^\alpha_\beta
       = {1 \over 2 a^3} \left[ a^2 \left( B^\alpha_{\;\;,\beta}
       + B_\beta^{\;\;|\alpha} \right) \right]^\cdot,
   \label{Rab} \\
   & & \delta R = 0, \quad 
       \delta ( R^a_b R^b_a ) = 0,
   \\
   & & R^{cd} R^0_{\;\; cd \alpha} = H^2 {\Delta \over a^2} B_\alpha,
   \nonumber \\
   & & \delta ( R^{cd} R^\alpha_{\;\; cd \beta} )
       = - {1 \over 2} \left( 3 \dot H + 2 H^2 \right)
       \left[ {1 \over a} \left( B^\alpha_{\;\;,\beta} 
       + B_\beta^{\;\;|\alpha} \right) \right]^\cdot
   \nonumber \\
   & & \qquad
       - {1 \over 2} \left( 7 \dot H + 6 H^2 \right) H {1 \over a}
       \left( B^\alpha_{\;\;,\beta} + B_\beta^{\;\;|\alpha} \right),
   \\
   & & R^{;0}_{\;\;\;\alpha} = 0, \quad
       \delta ( R^{;\alpha}_{\;\;\;\beta} )
       = - \dot R {1 \over 2a} \left( B^\alpha_{\;\;,\beta} 
       + B_\beta^{\;\;|\alpha} \right),
   \\
   & & \delta ( \Box R ) = 0,
   \\
   & & \Box R^0_\alpha = - \ddot R^0_\alpha - 5 H \dot R^0_\alpha
       + \left( 2 \dot H - 2 H^2 + {\Delta \over a^2} \right) R^0_\alpha,
   \nonumber \\
   & & \delta ( \Box R^\alpha_\beta )
       = - \delta \ddot R^\alpha_\beta - 3 H \delta \dot R^\alpha_\beta
       + \left( 2 H^2 + {\Delta \over a^2} \right) \delta R^\alpha_\beta
   \nonumber \\
   & & \qquad
       - H \left( 4 \dot H + {\Delta \over a^2} \right) {1 \over a}
       \left( B^\alpha_{\;\;,\beta} + B_\beta^{\;\;|\alpha} \right).
\eea


\end{document}